\documentclass{aa}
\usepackage{graphicx}
\usepackage{epsf,color}
\usepackage{natbib}
\usepackage{txfonts}
\definecolor{Mygreen}{rgb}{0.00, 0.72, 0.0}
\definecolor{Mypink}{rgb}{1.0, 0.0, 0.5}
\usepackage[breaklinks, citecolor=blue, linkcolor=blue, urlcolor=Mypink, colorlinks=true]{hyperref}

\bibpunct{(}{)}{;}{a}{}{,}

\usepackage{hyperref}

\def\apjl{ApJ}%
\def\apjs{ApJS}%
\def\aap{A\&A}%

\begin{document}

\title{Accurate sky signal reconstruction for ground-based spectroscopy with kinetic inductance detectors}
\author{A.~Fasano \inst{\ref{LAM}}\thanks{Corresponding author: Alessandro Fasano, \url{alessandro.fasano@lam.fr}}
        \and J.~F.~Mac\'ias-P\'erez\inst{\ref{LPSC}}
        \and A.~Benoit \inst{\ref{Neel}}
        \and M.~Aguiar \inst{\ref{IAC},\ref{laguna}}
        \and A.~Beelen\inst{\ref{LAM}}
        \and A.~Bideaud \inst{\ref{Neel}}
        \and J.~Bounmy\inst{\ref{LPSC}}
        \and O.~Bourrion \inst{\ref{LPSC}}
        \and G.~Bres \inst{\ref{Neel}}
        \and M.~Calvo \inst{\ref{Neel}}
        \and J.~A.~Castro-Almaz\'an \inst{\ref{IAC},\ref{laguna}}
        \and A.~Catalano\inst{\ref{LPSC}}
        \and P.~de~Bernardis\inst{\ref{Roma}}
        \and M.~De~Petris\inst{\ref{Roma}}
        \and A.~P.~de~Taoro \inst{\ref{IAC},\ref{laguna}}
        \and M.~Fern\'andez-Torreiro\inst{\ref{IAC},\ref{laguna}}
        \and G.~Garde \inst{\ref{Neel}}
    \and R.~G\'enova-Santos \inst{\ref{IAC},\ref{laguna}}
        \and A.~Gomez \inst{\ref{madrid}}
        \and M.~F. G\'omez-Renasco \inst{\ref{IAC},\ref{laguna}}
        \and J.~Goupy \inst{\ref{Neel}}
        \and C.~Hoarau\inst{\ref{LPSC}}
        \and R.~Hoyland \inst{\ref{IAC},\ref{laguna}} 
        \and G.~Lagache\inst{\ref{LAM}}
        \and J.~Marpaud\inst{\ref{LPSC}}
        \and M.~Marton\inst{\ref{LPSC}}
        \and A.~Monfardini \inst{\ref{Neel}}
        \and M.~W.~Peel \inst{\ref{IAC},\ref{laguna}}
        \and G.~Pisano \inst{\ref{Cardiff}}  
        \and N.~Ponthieu\inst{\ref{IPAG}}
        \and R.~Rebolo \inst{\ref{IAC},\ref{laguna}}
        \and S.~Roudier\inst{\ref{LPSC}}
        \and J.~A.~Rubi\~no-Mart\'in\inst{\ref{IAC},\ref{laguna}}
        \and D.~Tourres\inst{\ref{LPSC}} 
        \and C.~Tucker \inst{\ref{Cardiff}}
        \and C.~Vescovi \inst{\ref{LPSC}}
}

\institute{
    Aix Marseille Univ, CNRS, CNES, LAM, Marseille, France      \label{LAM}
        \and
    Univ. Grenoble Alpes, CNRS, LPSC/IN2P3, 38000 Grenoble, France
        \label{LPSC}
        \and
        Univ. Grenoble Alpes, CNRS, Grenoble INP, Institut Néel, 38000 Grenoble, France
        \label{Neel}
        \and
        Instituto de Astrofísica de Canarias, C/Vía Láctea, E-38205 La Laguna, Tenerife, Spain
        \label{IAC}
        \and
        Universidad de La Laguna, Dept. Astrofísica, E-38206 La Laguna, Tenerife, Spain
        \label{laguna}
        \and
        Dipartimento di Fisica, Sapienza Universit\`a di Roma, Piazzale Aldo Moro 5, I-00185 Roma, Italy
        \label{Roma}
        \and
    Centro de Astrobiología (CSIC-INTA), Torrejón de Ardoz, E-28850, Madrid, Spain 
        \label{madrid}
        \and
        Astronomy Instrumentation Group, University of Cardiff, The Parade CF24 3AA, UK
        \label{Cardiff}
        \and
        Univ. Grenoble Alpes, CNRS, IPAG, 38400 Saint Martin d'Hères, France 
        \label{IPAG}
}

\abstract
{Wide-field spectrometers are needed to deal with current astrophysical challenges that require multiband observations at millimeter wavelengths. An example of these is the KIDs Interferometer Spectrum Survey (KISS), which uses two arrays of kinetic inductance detectors (KIDs) coupled to a Martin-Puplett interferometer (MPI). KISS has a wide instantaneous field of view (1\,deg in diameter) and a spectral resolution of up to 1.45\,GHz in the 120--180\,GHz electromagnetic band. The instrument is installed on the 2.25\,m Q-U-I JOint TEnerife telescope at the Teide Observatory (Tenerife, Canary Islands), at an altitude of 2\,395\,m above sea level.}
{This work presents an original readout modulation method developed to improve the sky signal reconstruction accuracy for types of instruments for which a fast sampling frequency is required, both to remove atmospheric fluctuations and to perform full spectroscopic measurements on each sampled sky position.}
{We first demonstrate the feasibility of this technique using simulations. We then apply such a scheme to on-sky calibration.}
{We show that the sky signal can be reconstructed to better than 0.5\,\% for astrophysical sources, and to better than 2\,\% for large background variations such as in ``skydip'', in an ideal noiseless scenario. The readout modulation method is validated by observations on-sky during the KISS commissioning campaign.}
{We conclude that accurate photometry can be obtained for future KID-based interferometry using the MPI.}


\titlerunning{ Accurate sky signal reconstruction for spectroscopy with KIDs }
\authorrunning{A.~Fasano et al.} 

\maketitle

\section{Introduction}
Forthcoming scientific challenges in millimeter (mm) astronomy require large-scale spectroscopic mapping of the sky both to discriminate among the different components of the sky signal and to achieve
high sensitivity over large sky areas. This is particularly the case for large cosmological surveys aimed, for example, at detecting: the cosmic microwave background (CMB) primordial B-mode polarization signal and CMB spectral distortions (see e.g., \citealp{2019arXiv190901593C}); the overall matter distribution via lensing of the CMB photons and clusters of galaxies via the Sunyaev-Zel'dovich effect on the CMB   \citep[e.g.,][]{2020SPIE11443E..2FH,2021AAS...23721402H};  the cosmic infrared background \citep[CIB; e.g.,][]{2014A&A...571A..30P}; or the matter distribution via intensity mapping \citep[e.g.,][]{2020A&A...642A..60C}.\\

The natural choice for spectroscopic mapping is to use multichannel photometers, either on-chip (see e.g., \citealp{2018JLTP..193..170T}) or grating (see e.g., \citealp{2014SPIE.9153E..1WC}), because they minimize the noise while  simultaneously measuring different frequencies, as the optical load is divided over the different channels, minimizing the photon noise. The drawbacks of such techniques are the conception of large arrays dedicated to a single frequency bin, which massively increase the total pixel number, or a reduction of the field of view (FoV) adopting different techniques.
A more suitable candidate for the observation of extended and diffuse sources is represented by Fourier transform spectroscopy (FTS), which permits a wide instantaneous FoV (several\,degrees in diameter), exploiting a single large array for a wide range of frequencies.

Fourier transform spectroscopy relies on Fourier analysis by interfering two beams rather than separating the different light components through interference patterns (e.g., grating  or Fabry-Perot) or dispersive elements (prism or dichroic filters). This technique produces interference figures called interferograms that are analyzed by Fourier transform to retrieve the observed electromagnetic spectrum. 
Several state-of-the-art experiments at mm wavelengths for cosmology and extragalactic astrophysics have adopted such a spectrometric setup, starting from the first in space that paved the way, the Far-InfraRed Absolute Spectrophotometer (FIRAS) on board the COsmic Background Explorer satellite (COBE; \citealp{1993SPIE.2019..168M,1994ApJ...420..457F}), followed by the more recent Osservatorio nel Lontano Infrarosso Montato su Pallone Orientabile (OLIMPO; \citealp{2019JCAP...07..003M}), the KIDs Interferometer Spectrum Survey (KISS; \citealp{fasano-ltd}) and the CarbON CII line in post-rEionization and ReionizaTiOn epoch project (CONCERTO; \citealp{2020A&A...642A..60C}).

Sensitive detector arrays are required to properly sample the desired FoV and guarantee a high filling factor. In addition, the FTS capability to acquire the interferometric pattern of the astronomical sources in fractions of seconds is a crucial aspect, especially for ground-based experiments: removing atmospheric fluctuations requires fast acquisition to properly calibrate the signal and avoid signal contamination. In summary, the ground-based mm-wavelength experiments require the exploitation of low time-constant detectors, with high filling factor and fast interferometric pattern integration. To meet this requirement, we have developed a spectrum imager named KISS (for the instrument description see \citealp{fasano-nika2,fasano-ltd}), 
whose primary objective is
to act as a technological test bed for the CONCERTO instrument. KISS is a fast spectrometer whose detectors are based on kinetic inductance detector (KID) technology (\citealp{2003Natur.425..817D}) and is installed on one of the two 2.25\,m crossed-Dragone  Q-U-I JOint TEnerife (QUIJOTE; for a summary of the telescope see \citealp{quijote}) telescopes at the Teide Observatory (Tenerife, Canary Islands), at an altitude of 2\,395\,m above sea level (asl) in the northern hemisphere. The QUIJOTE telescopes are originally dedicated to characterizing the polarization of the CMB and other processes of Galactic and extragalactic emission in the frequency range 10--40\,GHz at large angular scales (see \citealp{2010ASSP...14..127R}).
Although the QUIJOTE instruments operate at frequencies below 40\,GHz, the optics of both telescopes are designed
to be used up to 200\,GHz, as required for KISS.

A critical aspect in the exploitation of KID arrays for scientific quality observations is the necessity to linearly convert the KID readout raw data into KID resonance frequency, which, in turn, is linear with the input sky power. For previous KID experiments such as Néel IRAM KID Array (NIKA; \citep{Monfardini_2010:NIKA} and New IRAM KID Array 2 (NIKA2; \citep{2018A&A...609A.115A}), a two-sample modulation readout scheme was developed to obtain sufficient photometric accuracy \citep{Calvo2013}. However, such a scheme is not applicable for fast sampling rates such as those required for fast FTS imaging instruments such as KISS and CONCERTO. Indeed, it would double the number of samples for equivalent readout performance and induce distortions in the interferogram shape leading to systematic errors in the reconstructed spectra.\\

In this paper, we address the challenge to extend the modulation technique for fast sampling rates while conserving equivalent photometric accuracy.
In Section~\ref{sec:photo}, we first explain the method used for NIKA (\citealp{Monfardini_2010:NIKA}) and NIKA2 (\citealp{2018A&A...609A.115A}) to convert the $(I,Q)$ data to resonance frequencies, and then we describe the new method for KISS, the data implementation, and the algorithm.
Section~\ref{sec:raw2science} presents our study of the feasibility of the KISS conversion technique using a physical model.
Finally, Sect.~\ref{sec:results} presents our results: the application of the method to real data for the characterization of the atmosphere and for photometry of Venus.

\section{From raw data to kinetic inductance detector resonance frequency shift}
\label{sec:photo}

\subsection{Kinetic inductance detector response and the 2-point modulation readout scheme}
{Kinetic inductance detectors are superconducting resonators that are sensitive to incoming light through the change of their resonance frequency, $f_{0}$. In practice, photons with sufficient energy to break Cooper pairs will modify the detector kinetic inductance and induce a frequency shift, $\delta f_{0}$, which is linearly proportional to the input optical power, $P_{\rm{opt}}$ \citep{Swenson}.
KIDs have been described in detail in several works (see \citealp{kids} for an overview) and are exploited in astronomical instruments: some examples are DEep Spectroscopic HIgh-redshift MAppe (DESHIMA; \citealp{DESHIMA}), NIKA and NIKA2 (\citealp{2016arXiv160508628C}), A-Microwave Kinetic Inductance Detector (A-MKID; \citealp{AMKID}), SuperSpec (\citealp{SuperSpec}), TolTEC\footnote{``TolTEC'' is not an acronym, but the name adopted for the instrument in honor of the Toltec ancient civilization of what is now central Mexico, it is where TolTEC will be ultimately installed.} (\citealp{TolTEC}), OLIMPO (\citealp{2019JPhCS1182a2005P}), MUltiwavelength Submillimeter kinetic Inductance Camera (MUSIC; (\citealp{music}), and the Ground-based B-mode Imaging Radiation Detector (GroundBIRD; \citealp{2020SPIE11445E..7QH}). \\

Kinetic inductance detectors are generally coupled to a microwave transmission line in which readout tone signals close to their resonant frequency are injected \citep{Swenson, bourrion2012}. The amplitude and phase of each transmitted and reflected signal are affected by the corresponding KID and are used to monitor the KID resonance frequency. As described in detail in Appendix~\ref{sec:kid}, the KID response can be well represented by a transfer function (ratio between the input and output tone signal) as a function of frequency, which can be approximated by a circle in the electrical In-phase and Quadrature $(I,Q)$ plane (see Fig~\ref{fig:IQ_modulation}). In the case of a single frequency tone per KID, the shift in resonance frequency can be reconstructed from the measured transmission for that tone. The first estimate of this can be obtained from the phase $\phi = \arctan \left( \frac{I}{Q} \right)$, which for small optical power variations will be linear with the KID resonance frequency shift. However, for bright sky signals and large background variations, as might be expected for ground-based experiments, the phase is not linear and this can lead to significant bias in the measurement of the input flux. To solve this problem, \citet{Calvo2013} developed the ``2-point'' modulation scheme \citep[see][for technical details on the electronics for the modulation]{bourrion2012} where the frequency of the input tone, $f_{\rm{tone}}$, is constantly radio-frequency (RF) modulated by the readout: $f_{\rm{tone}} \pm \Delta f$. Here, $f_{\rm{tone}}$ is the central tone frequency used to monitor the KID resonance and $\Delta f$ is the modulation in frequency  introduced. This enables instantaneous calibration of the measured transfer function variations in terms of frequency variations and then allows the user to define an integrated quantity that is linearly proportional to the input optical power.\\

The 2-point modulation scheme has proved to be very well adapted to observations with NIKA and NIKA2 in a large range of atmospheric conditions and for a large variety of astrophysical sources (\citep{Calvo2013}). However, it presents several drawbacks \citealp{2020A&A...637A..71P}: (1) it doubles the number of samples for an equivalent on-sky sampling rate, (2) the estimate of the variation of the KID transfer function corresponding to the modulation is noisy and needs to be smeared out (typically by a factor of 50 for NIKA and NIKA2), and (3) it is not well suited to fast variations of the input signal. As a consequence, the typical sampling rate for NIKA2 observations is 23.84\,Hz (47.68\,Hz in polarization mode), which is to be compared to the original electronic sampling rate of 1\,kHz.
}

\subsection{A 3-point modulation technique}
\label{subsec:3-point}

In the case of an FTS, the sky-sampling rate and the data sampling rate are expected to be very different: for each sky position, at least one interferogram (equivalent to one electromagnetic spectrum) needs to be obtained. The sky-sampling rate needs to be fast with respect to the scanning speed and with respect to the atmospheric fluctuations and drifts (both are affected by the same atmospheric emission). For KISS, which is a Martin-Puplett interferometer (MPI) (\citealp{fasano-ltd}), we acquire two interferograms: ``forward'' and ``backward''. They are obtained by introducing a phase delay in the optical path, which is called the optical path difference (OPD), by moving a mirror around the zero-phase delay, referred to as the zero optical path difference (ZPD). Considering atmospheric variations on timescales of greater than 1\,s, as suggested by atmospheric noise spectra at the IRAM 30\,m telescope \citep[see][]{2017A&A...599A..34R}, the sky-sampling
rate for KISS was safely set to 3.72\,Hz (0.268\,s) and the readout acquisition rate to 3.816\,kHz (262\,$\mu$s) to guarantee a large number of samples per interferogram so that the ZPD is properly sampled. Under these sampling rate conditions and taking into account that, near the ZPD we expect fast variations of the signal, the 2-point modulation technique cannot be applied either to KISS or CONCERTO. To overcome this issue, we developed a three-point (or 3-point) modulation scheme. \\

\begin{figure}[ht]
        \centering
        \includegraphics[width=.5\textwidth]{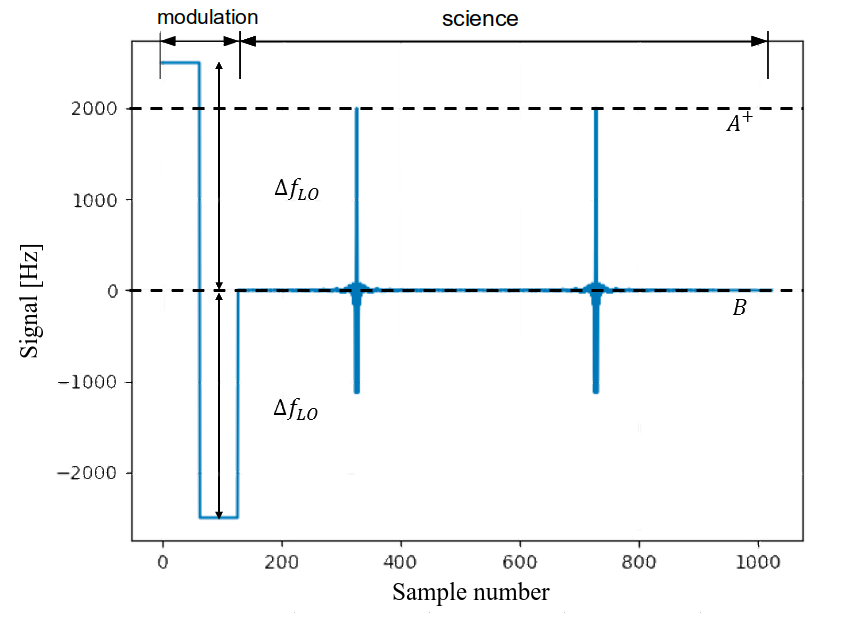}
        \caption{Expected KID resonance frequency, $\mathcal{I}$ (defined in Eq.~\ref{eq:freq_timeline}), as a function of the sample number within a KISS data bloc. Each data block consists of 1024 samples: the first 128 are dedicated to the modulation, with the remaining samples being dedicated to scientific data acquisition. $A^+$ identifies the ``interferometric response'' in the transmission output (see Sect.~\ref{subsec:phot}), while the ``background response'', $B$, (see Sect.~\ref{subsec:totpow}) is scaled to 0\,Hz; i.e., we subtracted $f_{\rm{tone}}$.}
        \label{fig:mod}
\end{figure}

On each sky position, we start by modulating the signal and then we acquire the two interferograms with no modulation. 
For this purpose, the KISS acquisition is organized in data blocks of 1024 samples (acquired at 3.816\,kHz) that correspond to a single sky position (acquired on-the-fly at  3.72\,Hz, while the telescope slews).  We show a synthetic data block in Fig.~\ref{fig:mod}. In the first (second) 64 samples, a positive (negative) RF bias modulation $f_{\rm{tone}}+\Delta f_{\rm{LO}}$ $\left( f_{\rm{tone}}-\Delta f_{\rm{LO}} \right)$ is applied, and then the remaining 896 samples are acquired with no modulation. 
In summary, the modulation with respect to the central $f_{\rm{tone}}$ can be written as
\begin{equation}
M(j)=   
        \begin{cases}
                +\Delta f_{ \mathrm{LO}} & 0 \leq j < 64 \\
                -\Delta f_{ \mathrm{LO}} &  64 \leq j < 128 \\
                 0 & 128 \leq j < 1024 \text{,} \\
        \end{cases}
\end{equation}
\noindent where $j$ represents the sample number.

 Overall, 87.5\,\% of the time is dedicated to scientific data acquisition without the injection of RF bias modulation and 12.5\,\% to the modulation for calibration purposes.
{This configuration fulfills the requirements imposed in terms of atmospheric noise fluctuations. Furthermore, with 448 samples per interferogram, it also achieves the scientific requirements in terms of spectral resolution and frequency coverage. Indeed, for a maximum spectral resolution of $\delta \nu = 1.45 $\,GHz we can retrieve the sky emission spectrum up to 650\,GHz, which is well beyond the typical maximum frequency for mm observations with ground-based experiments.}

The choice of the amplitude of the modulation results from a compromise between the noise level (modulation should be large enough to be easily detectable) and the accuracy of linear regime approximation, which is improved by adopting a modulation similar in amplitude to the input signal itself. \\

The typical spacing between KID resonance frequency has been accurately designed and has been measured in laboratory tests, resulting in a typical mean value of $1.3$\,MHz. This result was obtained by measuring the KID resonance frequency with a vector network analyzer in dark conditions. This value is large with respect to the frequency modulation which can be as large as a few tens of kilohertz. Furthermore, the relatively large spacing avoids interference between KIDs with adjacent resonances. Within these conditions, multiplexing factors of 400 can be obtained using a single electronic board of 500 MHz bandwidth \citep[e.g., ][]{bourrion2016}. For example, in the case of KISS, two electronic boxes are used to read 632 total pixels \citep{fasano-ltd} and for CONCERTO, 12 boxes are used to read 2 arrays of 2\,152 pixels each \citep{2020A&A...642A..60C}. Time synchronization between electronic boxes is guaranteed using a rubidium standard at 10\,MHz.

\subsection{Converting into KID resonance frequency shift}
\label{subsec:conv}
{
The KISS raw data are composed of $(I,Q)$ time-ordered data streams for each detector.
For a given data block and for each KID, we can define three points in the $(I,Q)$ plane, $p_1$, $p_2$, and $p_3$, by averaging the $I$ and $Q$ data on the positive modulation, negative modulation, and science regions, respectively, as shown in Fig.~\ref{fig:IQ_modulation} in arbitrary units (a.u.).

\begin{figure}[ht]
        \centering
        \includegraphics[width=0.45\textwidth]{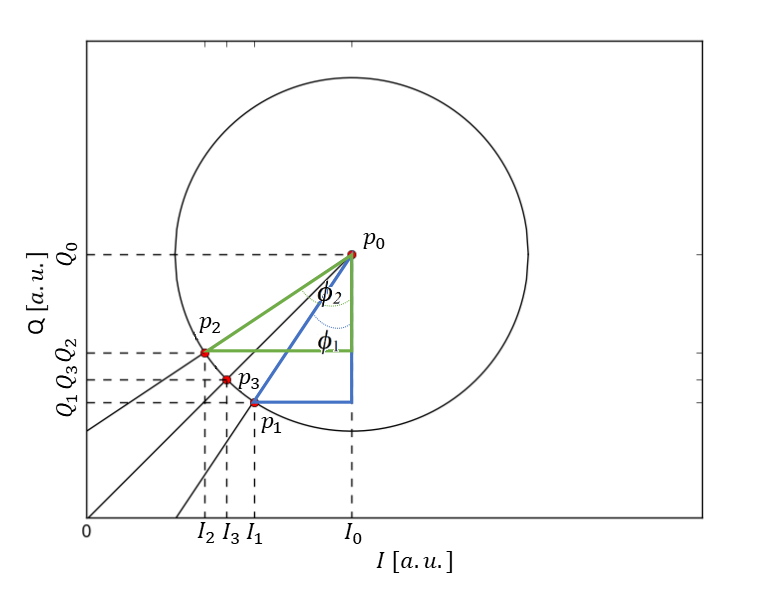}
        \caption{Resonance circle in the $(I,Q)$ plane in a.u.. $p_1=(I_1,Q_1)$ and $p_2=(I_2,Q_2)$ are the modulation points, while $p_3=(I_3,Q_3)$ is the measurement point. $p_0=(I_0,Q_0)$ is the circle center. }
        \label{fig:IQ_modulation}
\end{figure}

\noindent Furthermore, for each block, we fit a circle to this set of three points (defined in Fig.~\ref{fig:IQ_modulation}) and compute the circle center $p_0$ and radius $r_0$.
Thus, we can calculate the phases for $p_1$ and $p_2$
 with respect to $p_0$  : 

\begin{equation}
        \begin{split}
                \phi_1 &= \arctan\left( \frac{I_0-I_1}{Q_0 - Q_1}  \right) \text{,} & \\
                \phi_2 &= \arctan\left( \frac{I_0-I_2}{Q_0 - Q_2}  \right) \text{,} &      \end{split}
\end{equation}
\noindent from which a conversion factor, $C$, can be derived as

\begin{equation}
                C =  \frac{2\Delta f_{ \mathrm{LO}}}{\Delta\phi} \text{,}
\end{equation}
\noindent with $\Delta \phi = \phi_2-\phi_1$.

This conversion factor can be used to convert the scientific data into estimates of the KID resonance frequency shift. On one hand, we obtain a background value per block (per sky sample) as
\begin{equation}
    \Delta B = \left( \phi_3  - \phi_0 \right) \ C = \delta \phi \ C \text{,}
    \label{eq:freq_background}
\end{equation}
where  $\phi_3 = \arctan \left( \frac{I_0 - I_3}{Q_0 - Q_3} \right)$ and $\phi_0 = \arctan \left( \frac{I_0}{Q_0} \right)$.

\noindent On the other hand, we obtain a time-ordered interferogram signal as
\begin{equation}
        \mathcal{I}(t) = \left( \phi(t) - \phi_3 \right) \ C \text{,}
        \label{eq:freq_timeline}
\end{equation}
where $\phi(t) = \arctan \left( \frac{I_0 - I(t)}{Q_0 - Q(t)} \right)$. We note that $C$, $\Delta B,$ and $\mathcal{I}(t)$ are computed for each block and for each detector.\\

In the case of very large background fluctuations, the above procedure does not produce reliable results and we have extended it as follows:
\begin{equation}
        \label{eq:incre_calib}
        \Delta B_\mathrm{n} = \Delta B_\mathrm{n-1} + ( \delta \phi_\mathrm{n} - \delta \phi_\mathrm{n-1} ) \ C_\mathrm{n}  \text{,}
\end{equation}
\noindent where the $\mathrm{n}$ and $\mathrm{n-1}$ subscripts refer to two consecutive samples in time.
This corresponds to an ``incremental method'' that follows the evolution of the resonance circle, calculating a correction with respect to the previous $(I,Q)$ point. This complementary method is not adequate for fast and discontinuous signal variations but is fundamental for a specific observational case, which we discuss in Sect.~\ref{subsec:totpow} and show in Fig.~\ref{fig:old_incremental}. Nevertheless, for general purpose, it gives an equivalent performance to the normal conversion.
}

\section{Validation on simulations}
\label{sec:raw2science}

{
In this section, we validate the conversion procedure described above using simulations of the expected sky signal for the KISS instrument. To simulate the KISS detector response we use the KID properties described in Sect.~\ref{sec:kid}. We concentrate on the two most challenging aspects of the reconstruction of the sky
signal: determination of (1) the atmospheric opacity via skydip techniques and (2) the sky emission of the interferogram peak. For the former, we expect large background variations, while for the latter we expect fast temporal variations of the sky signal. 

\subsection{Sky signal model and simulations}
\label{sec:datamodelsim}
In the case of an MPI instrument, for instance, KISS and CONCERTO, we have two input sources (see \citealt{fasano-ltd}) that will interfere and produce two output components for which the intensity can be written as
}
\begin{equation}
        I^\pm(\delta) = \frac{1}{2}\left(E_1^2+E_2^2 \right) \pm \frac{1}{2}\left(E_1^2-E_2^2 \right)\cos(\delta) \text{,}
        \label{eq:MPI}
\end{equation}

where \noindent $E_1$ and $E_2$ are the electromagnetic components of the two input sources named ``sky'' and ``reference'' (the orthogonal components selected by the first MPI polarizer) and $\delta$ is the phase delay introduced by the OPD. The $\pm$ sign identifies the two (transmission-reflection) outputs, selected by the last MPI polarizer. \\

\noindent We can distinguish between two different approaches for the photometric exploitation of the interferograms (starting from Eq.~\ref{eq:MPI}), and we model both in this section. 
First, we can exploit the first term of Eq.~\ref{eq:MPI}, the sum of the two MPI input power sources, as:
\begin{equation}
    B =  \frac{1}{2} \left( E_1^2+E_2^2 \right) \text{.}
    \label{eq:baseline}
\end{equation}
We refer to this approach as the background response and we describe it in Sect.~\ref{subsec:totpow}.
This term will correspond to the background signal defined in Eq.~\ref{eq:freq_background}.
Secondly, we use ``interferometric response'' when we consider the interferogram amplitude: equivalent to the maximum and minimum (transmission and reflection output) signal at ZPD  (i.e., $\delta=0$), respectively:   

\begin{equation}
        A^\pm = \pm \frac{1}{2}\left(E_1^2-E_2^2 \right) \text{.}
\end{equation}
\noindent This approach is described in Sect.~\ref{subsec:phot}. \\

{
We use the average KID parameters for KISS that were measured in the laboratory (see Sect.~\ref{sec:kid}) to generate realistic simulations of the measured sky signal. The values of the parameters listed in Table~\ref{tab:sky_simu} are taken from laboratory characterization (Fig.~\ref{fig:hist}) discussed in Sect.~\ref{sec:kid}. The internal quality factor ($Q_\mathrm{i}$) is obtained with a ``sky'' background, that is, a 270\,K black body with a 20\% emissivity. Using Eq.~\ref{eq:S21},
we can convert the simulated signal into KID resonance frequency shift and derive $(I,Q)$. We generate data blocks for the different input signals considered and then we apply the conversion procedures described in Eqs.~\ref{eq:freq_background} and \ref{eq:freq_timeline}.}

\begin{table*}[ht]
        \footnotesize
        \centering
        \caption{Input values of the KID model. }
        \begin{tabular}{ccccccccccc}
        \hline \hline
                $\alpha$ & $t_0$ & $\phi_0$ & $Q_\mathrm{i}$ @ 50\,K flat field & $Q_\mathrm{c}$ & $f_0$ [MHz] & $\Delta f_{ \mathrm{LO}}$ [Hz] & $A^+$ [Hz] & $\tau$ & T$_\mathrm{atm}$ [K]  & Responsivity [Hz/K]  \\
        1 & 0 & 0 & 30\,000 & 21\,000 & 500 & 2\,500 & 500 & 0.15 & 270 & 500 \\ \hline
        \end{tabular}
        \label{tab:sky_simu}
\end{table*}

\subsection{Background response}
\label{subsec:totpow}

{

In this section, we concentrate on the accuracy of the reconstruction of the background signal. In the case of ground-based experiments such as KISS and CONCERTO, the main varying background signal comes from the atmospheric
emission. Furthermore, at mm wavelengths, the atmosphere absorbs the astrophysical signal, which needs to be corrected for prior to scientific exploitation. The atmospheric transmission and emission are related and can be determined \citep[see e.g.,][]{2009tra..book.....W} from the atmospheric opacity, $\tau$, which varies with frequency and pointing direction. The atmospheric opacity can be characterized by
performing elevation slews at constant azimuth called skydip \citep[see e.g.,][]{dragovan,archibald}.

In the case of KIDs, \citet{Catalano2014} showed that the measured resonance frequency as a function of air mass (am) is given by}

\begin{align}
        \label{eq:Skydip}
        F_\mathrm{Skydip}^\mathrm{Ground} &=    F_0 +\mathfrak{C} T_\mathrm{atm} \left[1- e^{ -\tau \cdot am } \right] \\ 
          & = F_0 + \Delta B(am) \text{,} \nonumber 
\end{align}

\noindent where $F_0$ is the instrumental offset corresponding to the frequency tone excitation for the considered detector for zero opacity, $\mathfrak{C}$ is the skydip conversion factor in Hz/K, $T_\mathrm{atm}$ (in Kelvin) is the temperature of the atmosphere (in particular the troposphere, below an altitude of 10 km, which is the dominant contributor), $am$ is the air mass, which in plane-parallel hypothesis is $ am \cong \frac{1}{\sin(el)}$ (see e.g.,\citealp{2009tra..book.....W} as a reference), and $\Delta B$ is the background response defined in Eq.~\ref{eq:baseline}, which follows the background evolution. The first term of Eq.~\ref{eq:Skydip} represents the skydip physical model and the second one the measured signal.  \\

We have developed a skydip model to validate the conversion technique and compare it with on-sky results.
In this model, we consider the de-focused FoV (referred to here as the ``de-focused sky'' and introduced in \citealp{fasano-ltd,2020A&A...642A..60C}) as the MPI reference source; the setup is thus sky on one MPI entrance and de-focused sky on the other. In such a configuration, the resultant interferogram is in principle negligible since there is no targeted source, $E_1 \approx E_2$ in Eq.~\ref{eq:MPI}. However, we observe a small residual interferogram that is the result of local opacity variation. In addition, the reference is coming from a larger portion of the sky (3\,deg in diameter) with respect to the KISS instantaneous FoV (1\,deg in diameter). It is therefore not possible to use the interferometric response for skydip observations and we are forced to exploit the background response. \\

Starting from the KID properties and the observational quantities reported in Table~\ref{tab:sky_simu}, we simulate skydip observations for a single detector. We use a high value of opacity of 0.15 at full bandwidth at the zenith, which represents the upper value exploited for real KISS observations, in order to evaluate the goodness of the method in the worst background conditions (i.e., highest frequency shift). In this simulation, the instrumental noise is not taken into account because it does not impact the evaluation of the systematic error that is induced by the large background variations.

\begin{figure}
        \centering
        \footnotesize
        \includegraphics[width=0.45\textwidth]{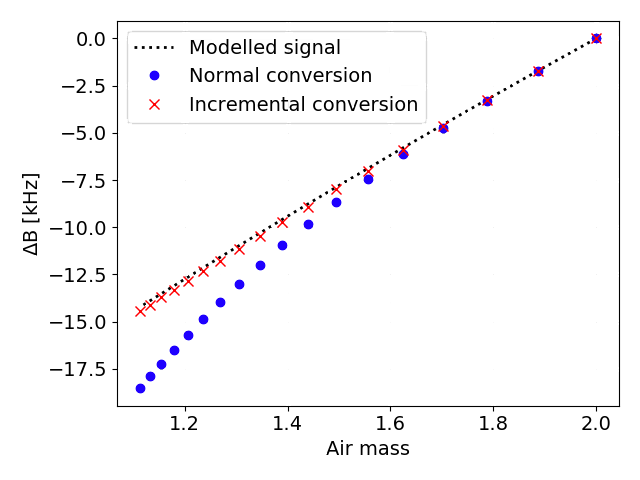}
        \caption{Background response, $\Delta B$, as a function of air mass, $am$, for the skydip model. In black: modeled input data. In blue: calculated data with normal conversion method. In red: calculated data with ``incremental method''. The signal is negative because the tuning is at high air mass ($am$=2) and the background signal diminishes at lower $am$.}
        \label{fig:old_incremental}
\end{figure}

{
An example of simulated skydip data is presented in Fig.~\ref{fig:old_incremental}, where we present the background signal variation as a function of air mass. The air mass variation corresponds to the typical elevation slew (from 30 to 65\,deg) performed for skydip observations with KISS.
The blue dots correspond to the background signal estimate in Eq.~\ref{eq:freq_background} and the red crosses to the incremental solution presented in Eq.~\ref{eq:incre_calib}. The simulated resonance frequency shift is shown as a black dotted line.
The major issue affecting the skydip observations is the large signal variation across the elevation slew. We observe that to reproduce the simulated data we need to use the incremental method. We computed that we can achieve a relative precision of 2\,\% at the maximum signal variation (at 1.1\,air masses) with a 2.5\,deg elevation sampling. This elevation sampling represents the best compromise between the noise reduction and the incremental method performance. \\}

\begin{figure*}[ht]
\centering
\includegraphics[width=.45\textwidth]{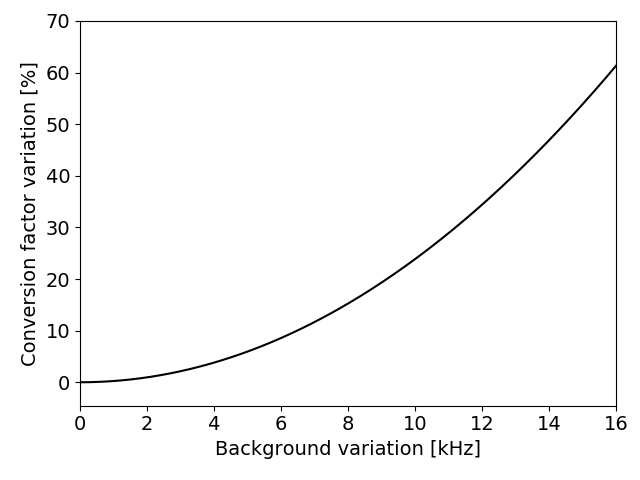}
\includegraphics[width=.45\textwidth]{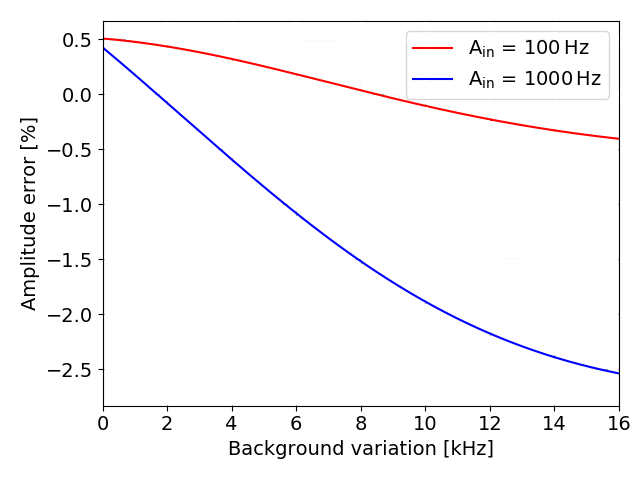}
\caption{Left: Conversion factor, $C$, variation as a function of background variation. Right: Bias in percentage in the reconstruction of the interferogram amplitude, $A$, as a function of background variation. The modulation factor is fixed at 2.5\,kHz. }
\label{fig:calamp_bck}
\end{figure*}

\subsection{Interferometric response}
\label{subsec:phot}

{
We concentrate now on the quality of the reconstruction of the interferogram amplitude signal and on the evolution of the conversion factor, $C$, with the background changes. Following the prescription in Sect.~\ref{sec:datamodelsim}, we have simulated KISS interferograms with different amplitudes and for various values of the sky background. We first generate raw data and then convert them into KID resonance frequency shifts using Eq.~\ref{eq:freq_timeline}. In the simulation, we assume the readout frequency tone is placed so that it corresponds to the KID resonance frequency for the minimum background considered. So, in the following, we can simply consider the variation with respect to this minimum background.} \\

In the left panel of Fig.~\ref{fig:calamp_bck}, we report the evolution of the conversion factor as a function of the background variation which is expressed in terms of background response, $\Delta B$. The conversion factor variation is expressed in percentage variation as $\left( \frac{C(t)-C_\mathrm{0}}{C_\mathrm{0}}\times 100 \right)$, where $C(t)$ is the conversion factor at a given background and $C_\mathrm{0}$ is the one at the initial background. We observe that the conversion coefficient increases with background variation.

\begin{figure}[ht]
        \centering
        \includegraphics[width=.45\textwidth]{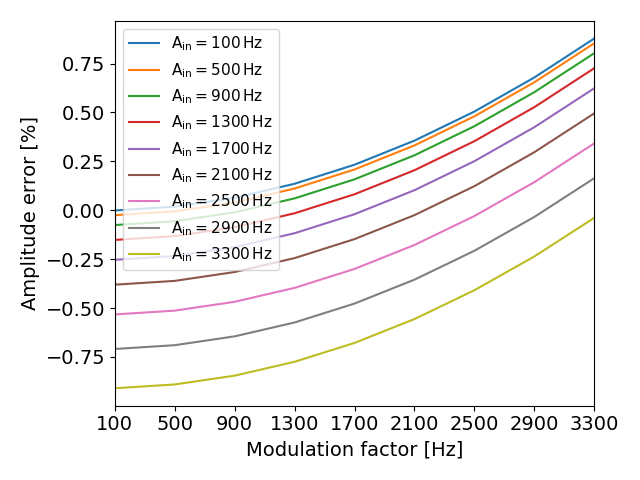}
        \caption{Systematic error in percentage variation of the interferogram amplitude as a function of modulation factor, $\Delta f_{ \mathrm{LO}}$.  The background is fixed at 0\,Hz. The error on the calculation converges to 0\,\% when the modulation factor approaches $A_\mathrm{in}$. }
        \label{fig:amp_mod}
\end{figure}

{
In the right panel of Fig.~\ref{fig:calamp_bck}, we report the relative bias in percentage in the reconstruction of the interferogram amplitude as a function of the background variation for a fixed modulation factor 2.5\,kHz. We computed the bias for two different input interferogram amplitudes: 100\,Hz (red solid curve) and 1\,kHz (blue solid curve).
We find that the relative bias increases exponentially for large background variations and is negative (we measure less flux than expected). 

However, we notice that for reasonable background variations (<5\,kHz), the relative bias on the signal is $\lesssim$1\%  both for small amplitude signals (100\,Hz) and large ones (1\,kHz).

\noindent In Fig.~\ref{fig:amp_mod} we investigate the dependency of the bias on the estimation of interferogram amplitude with the choice of the  modulation factor $\Delta f_{ \mathrm{LO}}$. For this, we have fixed the background variation to zero and have considered a wide range of input interferogram amplitudes, $A_\mathrm{in}$, ranging from 100 to 3\,300\,Hz. We observe that we converge to zero bias for modulation factors approaching $A_\mathrm{in}$. For KISS we have set a modulation factor of $\Delta f_{ \mathrm{LO}}=2.5$\,kHz to ensure a sufficient signal-to-noise ratio on the measurement of the conversion factor described above and to allow us to optimize the detectors between two sky observations (we set the readout frequency tone to the resonance frequency of the KID for the current background). Under these conditions, we expect a bias on $A_\mathrm{in}$ that is well below 0.5\,\% for most targets of interest. This calculation represents a lower limit as it does not take the noise into account. }

\section{On-sky results}
\label{sec:results}
\begin{figure}[ht]
        \centering
        \includegraphics[width=.50\textwidth]{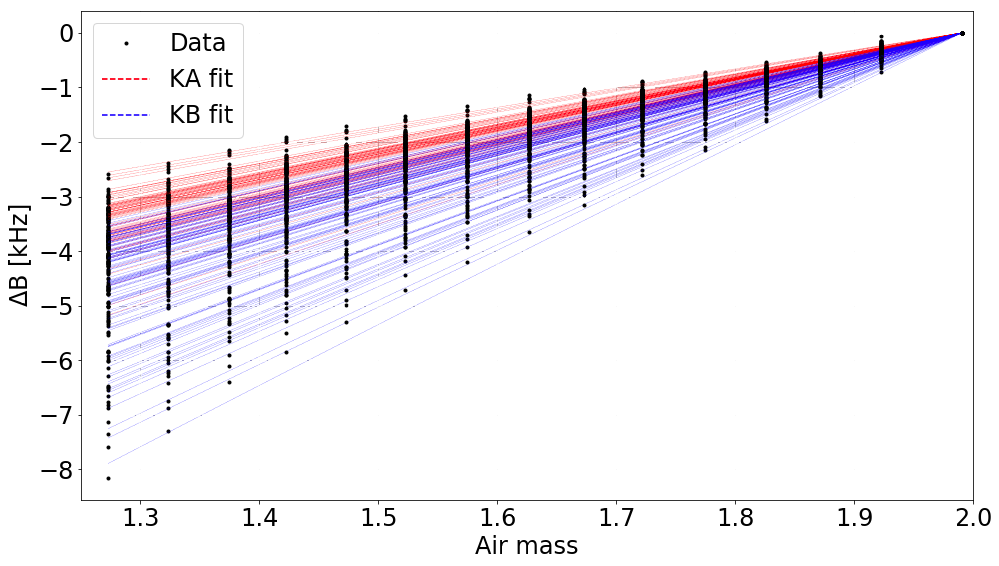}
        \caption{Background response, $\Delta B$, as a function of air mass. One example skydip for the best $\sim$42\% of the pixels, selected by a Chi-squared test. Black dots: binned signal. Red/blue dashed line: skydip fit distinguished per array. 
        The signal is negative because the tuning is at high air mass ($am$=2) and the background signal diminishes at lower $am$.}
        \label{fig:Skydip}
\end{figure}

{
\subsection{Determination of the integrated atmospheric opacity with skydip}

\subsubsection{Atmospheric opacity from satellite measurements of the  precipitable water vapor (PWV) }

Estimates of the atmospheric opacity at the Teide Observatory ---where the QUIJOTE telescope is installed--- can be obtained from measurements of the precipitable water vapor (PWV), and these can in turn be used to calculate the atmospheric opacity integrated on the KISS band. We use the atmospheric transmission at microwaves model (\texttt{ATM model})\footnote{\url{https://cab.inta-csic.es/users/jrpardo/atm.html}} to characterize the atmospheric contribution (see \citealp{pardo} for a description and \citealp{2001JQSRT..68..419P} for an application).\\

Precipitable water vapor data were obtained with the Global
Navigation Satellite System (GNSS) technique proposed by \cite{bev92,bev94}. The difference between the refracted and ideal straight-line paths followed by the signals coming from a set of satellites is perceived as delays by geodetic ground-based antennas. These delays can be estimated after a least-square fit of the signals (at 1.2\,GHz and 1.5\,GHz) received from a constellation of GNSS satellites ($\sim$10) averaging a time of around 2\,hours. The total delay, properly projected to the zenith and corrected for the ionospheric component, may be separated into two terms, the zenith hydrostatic delay (ZHD) and the zenith wet delay (ZWD) \citep{saa72}, which is directly proportional to the PWV.\\

The PWV values have been processed and calibrated by the IAC ``Sky Quality Team'' \citep{cas16} using data from the geodetic GNSS antenna named IZAN located at a distance of $\sim$1.5\,km from QUIJOTE telescopes. The antenna belongs to the Spanish Instituto Geogr\'afico Nacional, being part of The Regional Reference Frame Sub-Commission for Europe (EUREF) permanent network, from where the GNSS data are freely downloadable. The PWV values are retrieved every 30\,minutes.\\

As a reference, during the month of December 2019, we deduced atmospheric opacity values ---integrated over the KISS bandwidth--- of between 0.05 (very good) and 0.3 (not observable) with the \texttt{ATM model}, corresponding to a PWV ranging between 0 and 16\,mm, with a median value of 2.8\,mm.

\subsubsection{Skydip observations for atmospheric opacity determination}
We have regularly performed skydip observations during the commissioning of the KISS instrument at the QUIJOTE telescope, at different Azimuth angles. Each skydip observation consists of a continuous fast elevation slew during which KISS data are continuously acquired. This technique was preferred to the most common stepping elevation slew as it reduces the impact of atmospheric fluctuations during the measurements. At the beginning of the skydip observation, the KIDs are tuned so that the acquisition frequency tone lies exactly on the current KID resonance frequency (determined by the current background signal). For analysis purposes, we bin the data in intervals of 5\,deg in elevation to mitigate noise contributions. The raw data are then converted into resonance frequency shifts using the incremental method. This allows us to exploit a wide elevation range, 30--65\,deg (at higher elevation angles the air mass quickly converges to 1 and does not add information).
This wide elevation range is required to break the degeneracy between $\tau$ and $\mathfrak{C}$ parameters in the skydip equation (see Eq.~\ref{eq:Skydip}), to which the binned KISS data are fitted. The fit is performed iteratively for each KID in order to estimate $F_{0}$, $\tau,$ and $\mathfrak{C}$. In the first fitting iteration, we use the PWV as a prior to constrain the $\tau$ value and obtain $F_{0}$ and $\mathfrak{C}$. This procedure is applied for all the skydip observations available. The obtained $F_{0}$ and $\mathfrak{C}$ per skydip are averaged obtaining a set of new $F_{0}$ and $\mathfrak{C}$ parameters per pixel.
In the second iteration, we exploit the $F_{0}$ and $\mathfrak{C}$ obtained as a result of the first iteration to fit the $\tau$ per skydip observation.

We report in Fig.~\ref{fig:Skydip} an example of the measured skydip data for each KID (black dots). The best-fit model is represented as a solid colored line. The blue and red colors refer to the two MPI outputs that in the case of the KISS instrument are associated with two different KID arrays, named KA and KB, respectively.}

\subsubsection{Comparison to on-site PWV measurements}

\begin{figure}[ht]
        \centering
        \includegraphics[width=.45\textwidth]{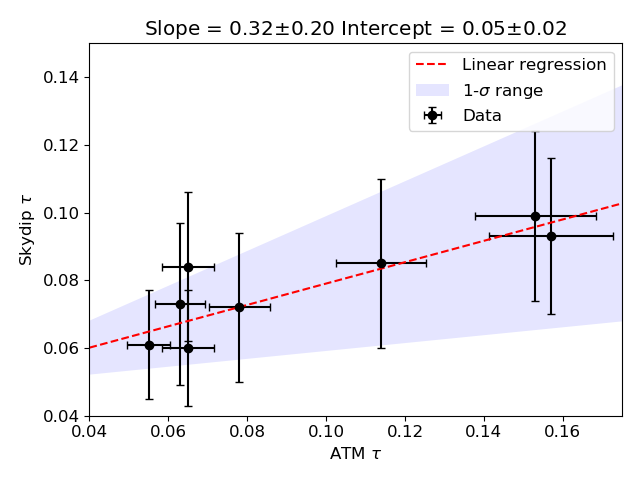}
        \caption{Zenith atmospheric opacity inferred by skydip as a function of the one derived by the \texttt{ATM model} from GNSS PWV data. The opacity values are integrated over the KISS bandwidth.
        The slope and the intercept of the linear regression are given above the figure. The error bars for the skydip are calculated on the statistics of the pixels, while the ones from the PWV method consider a 10\,\% conservative incertitude on the central value. In the red shaded region, we report the 1-$\sigma$ bounds of the slope, fixing the intercept at its central value. }
        \label{fig:linear}
\end{figure}

{
In Fig.~\ref{fig:linear}, we compare the atmospheric opacity values obtained from the skydip observations and those obtained from the on-site PWV estimates. These represent the opacity values integrated over the KISS band and taken to the zenith angle. We find that the two estimates are consistent as demonstrated by linear regression, for which we obtain $\mathrm{r}=0.86$ and $\mathrm{p}=6 \times 10^{-3}$.} 
On the other hand, for the higher opacities in Fig.~\ref{fig:linear} (corresponding to PWV$<$3\,mm) the two methods present a lower correlation. 
This is probably due to the incertitude on the PWV calculation coming from the calibration errors and the propagation of all the uncertainties in the PWV determination, of namely $\sim$30\,\% at 3\,mm for PWV.\footnote{$\Delta\text{PWV}=0.02\times \sqrt{\text{PWV
}^2 + 1575\text{ mm}^2}$, see \cite{pwv}}
In addition, KISS and the GNSS antennas do not observe exactly the same air column as they are $\sim$1.5\,km apart. Furthermore, the GNSS zenith PWV values are obtained after processing the signals of satellites with different positions on the sky, which are mapped to the zenith presuming azimuth symmetry. Therefore, even assuming a great horizontal homogeneity in the distribution of PWV, there may be some degrees of point-wise de-correlation in specific directions. All these factors can explain the observed discrepancy.
Finally, the method is improved each time a skydip measurement is performed, since the $\mathfrak{C}$ is better constrained by the increased statistics.

Exploiting the PWV as an initial value to break the parameters of the skydip method is a promising method for the opacity correction in KISS.

\subsection{Venus observations with KISS}
\begin{figure}[ht]
        \centering
        \includegraphics[width=.5\textwidth]{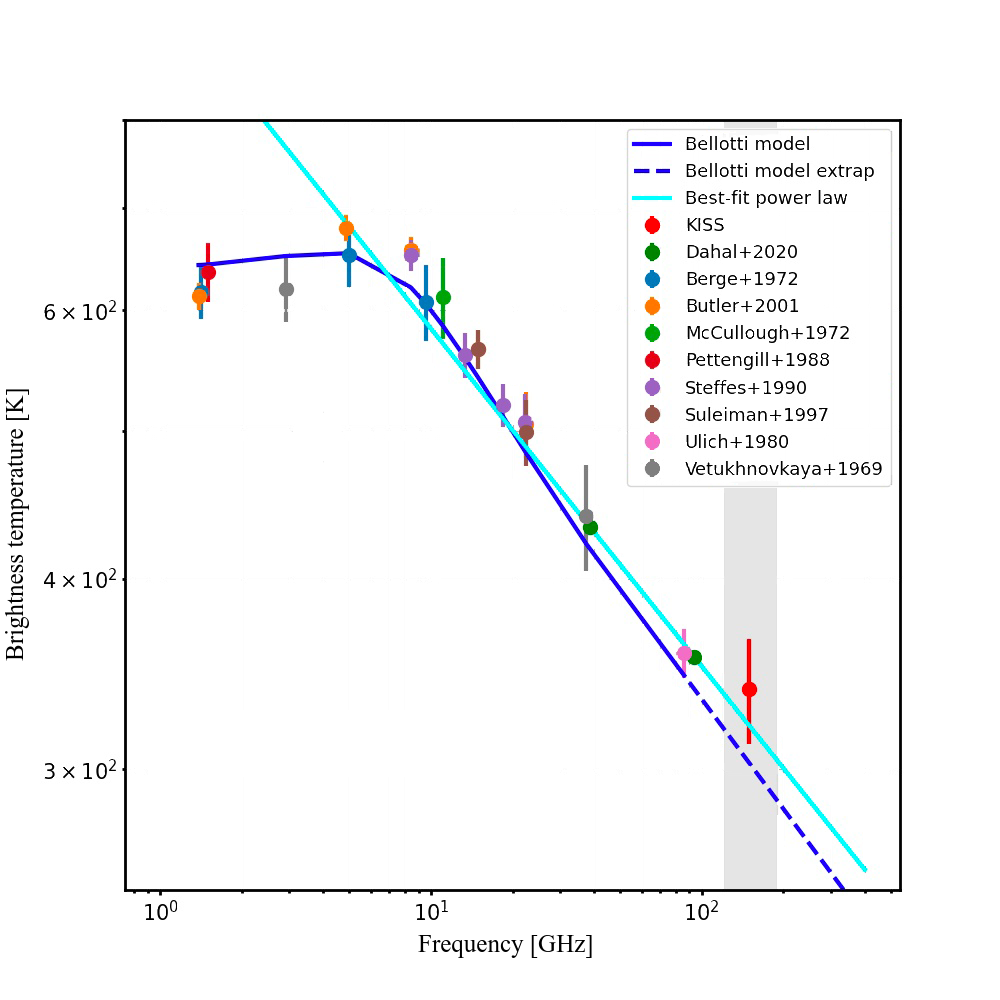}
        \caption{Venus brightness temperature as a function of frequency. We present a collection of radio and mm wavelength measurements including those of KISS.
        The blue solid line identifies the \citet{Bellotti:2015} model and the dashed one represents its power-law extrapolation. The cyan solid line identifies a single power-law.}
        \label{fig:venus}
\end{figure}

During the KISS commissioning campaign, we regularly observed the planets Jupiter and Venus, which can be considered as point sources. Using Jupiter as an absolute calibrator (its brightness temperature is known to about 5\,\% uncertainty) and the KISS data reduction pipeline, which will be described in \citet{kiss_pipe}, we estimated the Venus brightness temperature from the KISS observations considering only the background component. We obtain a brightness temperature of $T^{\rm{KISS}}_{b,\rm{Venus}} = 338 \pm 27$\,K  at 150\,GHz (the central frequency of the 120--180\,GHz KISS band). In Fig.~\ref{fig:venus} we present the KISS measured brightness temperature compared to measurements from other experiments \citep{Dahal:2020nfp, Berge:1972, Butler:2001,McCullough:1972,Pettengill:1988,Steffes:1990,Suleiman:1997,Ulich:1980,Vetukhnovkaya:1988} for frequencies ranging from 1 to 100\,GHz.  We observe that the \citet{Bellotti:2015} model underestimates the power at high frequency. A single power-law, $T_{b,\rm{Venus}}(\nu) = A  \times \left( \nu/\rm{1\,GHz}\right)^{\beta}$, is shown, which is used to fit the high-frequency data (above 20\,GHz) excluding the KISS measurement. The best-fit power-law parameters are $A = \left ( 972 \pm 1 \right)$\,K and $\beta = -2.73 \pm 0.01$. The KISS data are consistent with this power-law model. Thus, we conclude that the photometric calibration procedure described in this paper leads to accurate flux estimates with KISS limited mainly by absolute calibration.

\section{Conclusions}
\label{sec:conclu}
{
There is strong demand from the astrophysics and cosmology community to develop mm  wavelength wide-field multiband instruments. Multiband or spectroscopic observations would help for instance to discriminate among different astrophysical components, to characterize foreground emission, to measure CMB spectral distortions, and for line intensity mapping (see e.g., \citealp{2017arXiv170602464A}). 
Interesting instrumental candidates to fulfill such requests are FTS-based instruments made of large arrays of fast detectors, for instance KIDs. In the case of ground-based experiments, the latter is of primary importance as we expect a sampling rate of a few kHz to be able both to mitigate the impact of atmospheric fluctuations (on temporal scales of about 1\,Hz) and obtain full spectral coverage for each position on the sky (hundreds of samples).  \\

For KID-based instruments, the main observational challenge is to be able to convert the acquired raw data into KID resonance frequency, which is proportional to the input sky power. The standard conversion techniques developed for previous KIDs experiments such as NIKA and NIKA2 cannot be used for such high sampling rates.
Taking the KISS instrument installed on the QUIJOTE telescope at the Teide Observatory (Tenerife, Canary Islands)  as an example, we have demonstrated that this conversion is possible by using an innovative 3-point readout modulation technique. Using simulations, we show that the sky signal can be reconstructed to better than 2\,\% in the case of large sky background variations during skydip observations, and to better than 0.5\,\% for most astrophysical targets of interest. These values refer to noiseless simulations and they show that the method does not limit the signal reconstruction precision.
These results have been corroborated by on-sky observations during the commissioning of the KISS instrument both in the case of skydip and point-source observations. \\

This conversion procedure can be applied to any FTS instrument employing KIDs, paving the way to a new generation of wide-field spectrometers. This is the case for the CONCERTO instrument (\citealp{2020A&A...642A..60C})\footnote{ \url{https://mission.lam.fr/concerto/} } recently installed on the Atacama Pathfinder Experiment (APEX) telescope in Llano de Chajnantor (Chile), at an altitude of 5\,105\,m asl. }

\begin{acknowledgements}
The KID arrays described in this paper have been produced at the PTA Grenoble microfabrication facility. 
This work has been partially supported by the LabEx FOCUS ANR-11-LABX-0013, the European Research Council (ERC) under the European Union's Horizon 2020 research and innovation program (project CONCERTO, grant agreement No 788212), and the Excellence Initiative of Aix-Marseille University-A*Midex, a French ``Investissements d'Avenir'' program.
This work has been partially funded by the Spanish Ministry of Science under the project AYA2017-84185-P.
Based on observations made with the first QUIJOTE telescope (QT-1), operated on the island of Tenerife in the Spanish Observatorio del Teide of the Instituto de Astrofisica de Canarias.
We thank the Sky Quality Team at IAC to have supported analyzing the PWV data. 
AF thanks to the Université Grenoble Alpes for having funded his Ph.D., dedicated to the KISS experiment.

This research made use of {\tt Astropy} (\url{http://www.astropy.org}) a community-developed core Python package for Astronomy \citep{astropy:2013, astropy:2018}. We also use  {\tt Matplotlib} (\url{https://matplotlib.org}, \citep{Hunter:2007}), {\tt NumPy} (\url{https://numpy.org}, \citealp{harris2020array}) and {\tt SciPy} (\url{http://www.scipy.org}, \citealp{2020SciPy-NMeth}).
\end{acknowledgements}

\bibliographystyle{aa}
\bibliography{bibliography}

\appendix

\section{KID properties}
\label{sec:kid}

In this Appendix, we introduce the quantities required to describe the main characteristics of the detectors.
KIDs consist of a high-quality factor resonator ---cooled down to cryogenic temperatures to reach the superconducting regime--- coupled to a microwave transmission line that carries a readout tone signal. In the readout system, each KID is associated with an excitation tone, which corresponds to an estimate of its resonance frequency for a specific optical load.
The resonance frequency, $f_0$, is the frequency at which the resonator reflects the bias energy, acting as a band-stop filter. 
This effect is read out and recorded in a coupled feedline and is quantified by the coupling quality factor, $Q_\mathrm{c}$. Additionally, the resonator internal quality factor, $Q_\mathrm{i}$ quantifies the ratio of the fraction of energy that is lost in the AC cycle to the total energy stored in the resonator itself.
The physical distance between the pixel and the feedline is chosen to satisfy the optimal coupling conditions, which is achieved when $Q_\mathrm{c}$  is  of  the  same  order  as  $Q_\mathrm{i}$ at typical background loading. For NIKA2, such a value for  $Q_\mathrm{c}$   is optimized at a few thousands (\citealp{2018A&A...609A.115A}).

The quality factors are related by the equation of the resonator (or total) quality factor, $Q_{ \mathrm{res}}$, which represents the ratio of the energy fraction that is lost ($Q_\mathrm{i}$ dissipating and $Q_\mathrm{c}$ leaking) per cycle by the total resonator system, that is the pixel and the coupled bias line to the one stored:

\begin{equation}
        \label{eq:Q_tot}
        Q_{ \mathrm{res}} = \left( \frac{1}{Q_\mathrm{i}}+\frac{1}{Q_\mathrm{c}} \right)^{-1} \text{.}
\end{equation}

A KID measures the Cooper pair population change and the resulting frequency shift is converted to an input optical power (as demonstrated in \citealp{Swenson}).
Its time constant is fixed by the recombination time of the quasi-particles (a few tens of\,$\mu s$). This represents a major advantage with respect to the other detectors, which are a factor $\gtrsim$10 slower (e.g., thermal bolometers and transition-edge sensor, TES, \citealp{2020A&A...641A.179C}).

The bias signal described by the scatter parameter $S_{21}$ related to the single KID detector is studied in the complex plane $(I,Q)$. $S_{21}$ is the ratio of the output, $S_2$, to the input, $S_1$, signal. The subscripts  1 and 2
 denote the ``electrical ports'' for the feedline system, respectively the input injected port and the output readout one.

The model that describes the electrical characteristics of the single pixel is based on the $S_{21}$ definition (see \citealp{Gao} for a detailed description):

\begin{equation}
        S_{21}(f)= \alpha e^{-2\pi j f t_0} \left[ 1-\frac{\frac{Q_{ \mathrm{res}}}{Q_\mathrm{c}}e^{j\phi_0}}{1+2jQ_{ \mathrm{res}}\left(\frac{f-f_0}{f_0}\right)}\right] \text{,}
        \label{eq:S21}
\end{equation}

\noindent where $f$ is the bias frequency, $\alpha$ is a complex constant accounting for the gain and phase shift through the system, $e^{-2\pi j f\tau}$ corrects the cable delay of the readout system with a time $t_0,$ and $\phi_0$ is the resonance phase.

The conversion of the $(I,Q)$ signal to absorbed optical power is one of the most difficult challenges when using KIDs.
We exploit the laboratory measurement of $f_0$ as well as $Q_\mathrm{i}$ and $Q_\mathrm{c}$ to characterize the resonator. We use the skewed Lorentzian profile (defined in \citealp{Gao}) to fit the resonance amplitude:

\begin{equation}
    \left|S_{21}(f)\right|= a+b(f-f_0)+\frac{c+d(f-f_0)}{1+4Q_{ \mathrm{res}}^2\cfrac{(f-f_0)^2}{f_0^2}} \text{,}
\label{eq:s21_amp}
\end{equation}

\noindent where $a$, $b$, $c,$ and $d$ are factors that do not influence the parameters under study ($f_0$, $Q_\mathrm{i}$ and $Q_\mathrm{c}$). With Eq. \ref{eq:s21_amp} we can calculate $Q_\mathrm{i}$ (\citealp{Gao}) as

\begin{equation}
    Q_\mathrm{i} \approx \frac{Q_{ \mathrm{res}}}{\mathrm{min}(\left|S_{21}(f)\right|} \text{.}
\end{equation} 

\begin{figure}[ht]
        \centering
        \includegraphics[width=.5\textwidth]{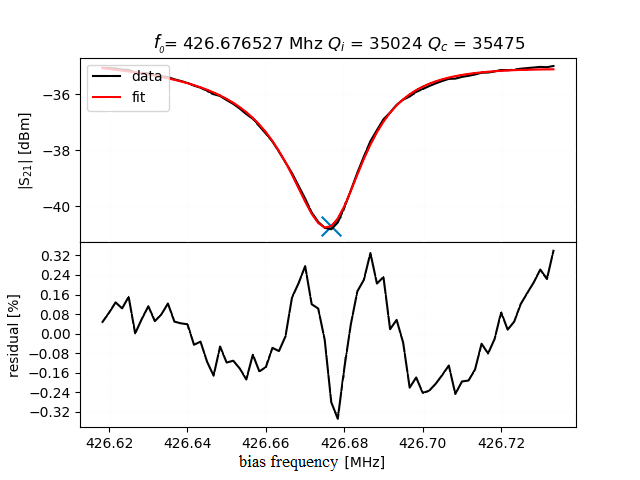}
        \caption{Amplitude signal, $\left|S_{21}(f)\right|$, as a function of the bias frequency, $f$. The resonance curve is fitted to characterize the pixel. Top: Single pixel bias signal (in black) and fitting Eq.~\ref{eq:s21_amp} (in red). The blue cross identifies the resonance frequency. Bottom: Percentage residual of a good fit between data and model.}
        \label{fig:fit_amp}
\end{figure}

\begin{figure}[ht]
        \centering
        \includegraphics[width=.5\textwidth]{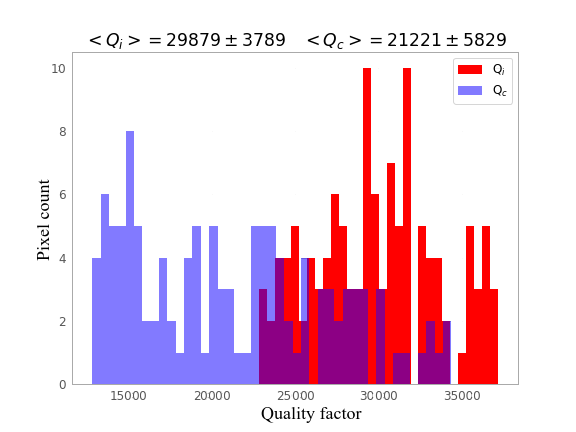}
        \caption{Quality-factor histograms measured on the A KISS array (for the best $\sim$30\,\% pixels). The $Q_\mathrm{i}$ measured with a flat-field background of 50\,K that models the sky is shown in red and  $Q_\mathrm{c}$ is shown in blue. The quality factors are represented in shadowed colors. The pixels are well distributed around suitable values, and the scatter is due to the different pixel designs that vary the resonance frequency and to slight fabrication defects.}
        \label{fig:hist}
\end{figure}

In Fig.~\ref{fig:fit_amp} and \ref{fig:hist} we show the electrical measurements for the KISS arrays performed in the laboratory. The figures show the standard electrical characterization with the fitted $f_0$, $Q_\mathrm{i}$, and $Q_\mathrm{c}$ values. This process represents the first step toward the full validation of the detectors. Furthermore, we exploit these quantities to compute a realistic model in Sect.~\ref{sec:raw2science}.

In Fig.~\ref{fig:fit_amp}, we report an example of a single pixel fitting. The procedure returns low residuals and gives the electrical characterization of the KID in terms of the $f_0$, $Q_\mathrm{i}$, and $Q_\mathrm{c}$ parameters.
In addition, all pixels of both arrays provide good fits with homogeneously distributed resonances, and have an average of $\sim$8\,dBm in depth with a few tens of kHz in width. Finally, the resonances are within the optimal working range for the amplifiers 400--900\,MHz.
These results are comparable to those of  NIKA2, which have already been demonstrated to provide good on-sky performance. In the KISS case, the $Q_\mathrm{i}$ is a factor of $\sim$2-3 smaller because the bandwidth is a factor two broader.

\noindent Figure~\ref{fig:hist} shows the similarity of $Q_\mathrm{c}$ and $Q_\mathrm{i}$ (compatible at 1 standard deviation). The detectors thus reach the so-called ``critical condition''. The quality factors overlap in the histogram with a dispersion compatible with the few thousand units requirement.
Higher $Q_\mathrm{c}$ would mean stronger coupling with the feedline, which translates to higher interaction between the pixel and the feedline, and low $Q_{ \mathrm{res}}$. This causes the resonance to be deeper and broader, in which case it is easier to identify. On the other hand, high $Q_{ \mathrm{res}}$ makes it easier to saturate the resonance for small optical loads.
$Q_\mathrm{i}$ is fixed by the electromagnetic bandwidth and the optical loading. The matching of the critical condition represents the compromise between the resonance depth (on the amplitude) and the high $Q_{ \mathrm{res}}$ optimized for the background.

\end{document}